\documentclass[aps,eqsecnum,nofootinbib,superscriptaddress]{revtex4}
\usepackage{amsmath}
\usepackage{comment}
\newcommand{\half}{ \frac{1}{2} }
\newcommand{\dotX}{\dot{X}}

\newcommand{\bra}[1]{\big\langle~#1~\big|}
\newcommand{\ket}[1]{\big|~#1~\big\rangle}

\newcommand{\Exp}[1]{\big\langle~#1~\big\rangle}
\begin{document}
%
\title{Hall Conductivity for Charged Strings
and Modified Spectrum-Generating Algebra}
\author{Akira Kokado}
\email{kokado@kobe-kiu.ac.jp}
\affiliation{Kobe International University, Kobe 658-0032, Japan}
\author{Gaku Konisi}
\email{konisi@wombat.zaq.ne.jp}
\affiliation{Department of Physics, Kwansei Gakuin University,
Sanda 669-1337, Japan}
\author{Takesi Saito}
\email{tsaito@k7.dion.ne.jp}
\affiliation{Department of Physics, Kwansei Gakuin University,
Sanda 669-1337, Japan}
\date{\today}
\begin{abstract}
When open strings end with a charge $q$ on a D2-brane,
which involves constant background magnetic field $B$
perpendicular to the brane,
we calculate the Hall conductivity for these charged strings.
We also construct the corresponding spectrum-generating algebra,
which assures that our system is ghost-free under some conditions.
\end{abstract}
\maketitle
\section{Introduction}
The Hall effect%
\cite{ThoulessEtAL}
concerns two-dimensional electron system,
which is supplied with a very thin flat conductor
placed in a constant uniform magnetic field $B$
perpendicular to the plate.
The Hall effect is characterized by the induced electric field $E_1$
in the $1$-direction on the plate,
and the induced current density $J_2 = \sigma_{21}\, E_1$
in the $2$-direction, where $\sigma_{21}$ is the Hall conductivity
given by 
\begin{align}
  & \sigma_{21} = - e\, n/B~,
\label{eq:def-class_Hall_cond}
\end{align}
$e\, n$ being the charge density.
Here there is no induced current in the $1$-direction,
i.e., $J_1 = 0$.

The quantum Hall effect is characterized by the formula
\begin{align}
  & \sigma_{21} = - e^2\, \nu/h~,
\label{eq:def-quant_Hall_cond}
\end{align}
where $\nu = n\, h/e\, B$ takes integer or fractional values
against $n$ or $B$
when the system is laid in the very strong magnetic field and
in the extremely low temperature.
It is known to come from effects of impurities
or electron-electron scatterings%
\cite{ThoulessEtAL}.

In this paper we are interested in the Hall effect
for the charged strings on the D2-brane.
However, we do not consider the quantum Hall effect,
but we confine ourselves to the ordinary pure Hall effect.
We consider an open string, $X^\mu(\tau, \sigma)$,~
$0 \le \sigma \le \pi$\,, which ends on the D2-brane at $\sigma = 0$,
but the other side $\sigma = \pi$ is not on the brane.
The string carries a charge $q$ located only
at the end point $\sigma = 0$.
The D2-brane involves a constant magnetic field $B$
perpendicular to the brane.

This system is clearly equivalent to the above Hall system,
if the charged strings are replaced by electrons
and the D2-brane by the flat thin conductor
placed in a constant uniform magnetic field $B$
perpendicular to the plate. 

Upon the quantization of the charged string, we meet a problem
of how to work the no-ghost theorem in our system.
For the free string the spectrum-generating algebra (SGA) is well
known to assure the ghost-free for string states if the space-time
dimension ($d$) is $d=26$ and the Regge intercept $\alpha (0)$ 
is $\alpha (0)=1$. For the charged string with a background field,
however, it is not clear whether the conventional SGA works well
as it is.
We then consider the modification of SGA,
and will find that our system is ghost-free
if $d=26$ and $\alpha (0)=1-\omega /2 +\omega ^2/2$,
where $\omega$ is the cyclotron frequency of the charged string.

In Sec.\ref{sec:qunantize}
we summarize the quantization of the charged string
placed in a constant electromagnetic field.
In Sec.\ref{sec:Hall_cond}
we calculate the Hall conductivity for the charged strings.
In Sec.\ref{sec:SGAstring} the modified spectrum-generating algebra
is constructed.
The final section is devoted to the concluding remarks.
\section{Quantization of the charged string} \label{sec:qunantize}
The interaction Lagrangian of our system is given by
\begin{align}
  & L_I
  = q\, \dotX_\mu(\tau, \sigma)\, A^\mu\big( X(\tau, \sigma) \big)~
      \Big\vert_{\sigma=0}
  = - q\, \half\, \dotX_\mu(\tau, \sigma=0)~F^{\mu}_{\ \nu}~
     X^\nu(\tau, \sigma=0)~.
\label{eq:def-L_int}
\end{align}
Here we have chosen a gauge such as
\begin{align}
  & A^\mu = -\frac{1}{2}F^{\mu}_{\ \nu}~X^\nu~,
\label{eq:gauge}
\end{align}
with $F^{\mu}_{\ \nu}$, the constant antisymmetric field strength
only for $\mu,\, \nu = 0, 1, 2$, i.e.,
\begin{align}
  & F^{\mu}_{\ \nu}
  = \begin{pmatrix} 0 &  E & 0 \\ E & 0 & B \\ 0 & - B & 0
    \end{pmatrix}~.
\label{eq:def-F}
\end{align}

Introducing a time-independent function $\rho(\sigma)$ such that
\begin{align}
  & \rho(0) = q~, & & \rho(\pi) = 0~,
\end{align}
the interaction Lagrangian can be rewritten as
\begin{align}
   L_I
  &= \half~\Big[~\rho(\sigma)\, \dotX_\mu(\tau, \sigma)\,
     F^{\mu}_{\ \nu}\, X^\nu(\tau, \sigma)~\Big]^\pi_0
  = \half \int^\pi_0 d\sigma~\partial_\sigma \Big(
     \rho(\sigma)\, \dotX_\mu(\tau, \sigma)\,
        F^{\mu}_{\ \nu}\, X^\nu(\tau, \sigma) \Big)
\nonumber \\
  &= \half \int^\pi_0 d\sigma~\Big( 
     \rho'\, \dotX_\mu\, F^{\mu}_{\ \nu}\, X^\nu
  + \rho\, \dotX_\mu\, F^{\mu}_{\ \nu}\, \big( X^\nu \big)'
  + \rho\, \big( \dotX_{\mu} \big)'\, F^{\mu}_{\ \nu}\, X^\nu \Big)
\nonumber \\
  &= \half \int^\pi_0 d\sigma~\Big(
     \rho'\, \dotX_\mu\, F^{\mu}_{\ \nu}\, X^\nu
  + \rho\, \dotX_\mu\, F^{\mu}_{\ \nu}\, \big( X^\nu \big)'
  - \rho\, \big( X_{\mu} \big)'\, F^{\mu}_{\ \nu}\, \dotX^\nu \Big)
\nonumber \\
  &= \half \int^\pi_0 d\sigma~\Big(
     \rho'\, \dotX_\mu\, F^{\mu}_{\ \nu}\, X^\nu
  + 2\, \rho\, \dotX_\mu\, F^{\mu}_{\ \nu}\, \big( X^\nu \big)' \Big)~,
\label{eq:def-L_int-II}
\end{align}
where the total time derivative $\partial_\tau\big( \rho X' F X \big)$
has been discarded.
This theory does not depend on the functional form of $\rho(\sigma)$
for $0 < \sigma < \pi$,
because the action based on Eq.(\ref{eq:def-L_int-II})
is invariant under a variation with respect to $\rho(\sigma)$.

The total Lagrangian is given as
\begin{align}
  & L
  = \frac{1}{4 \pi \alpha'} \int^\pi_0 d\sigma~
       \left[~\dotX_\mu\, \dotX^\mu
    - {X_\mu}' {X^\mu}'~\right]
  + \half \int^\pi_0 d\sigma~\dotX_\mu F^{\mu}_{\ \nu} \left[~
     \rho'\, X^\nu + 2 \rho\,  {X^\nu}'~\right]~.
\label{eq:def-L}
\end{align}
The equation of motion and the boundary conditions
follow from the action based on Eq.(\ref{eq:def-L}):
\begin{align}
  & \ddot{X}^\mu - \big( X^\mu \big)'' = 0~,
\label{eq:EOM} \\
  & \left[~\big( X^\mu \big)'
  + 2 \pi \alpha'\, \rho F^\mu{}_\nu \dotX^\nu~\right]\,
  \Big\vert_{\sigma=0, \pi} = 0~.
\label{eq:bc}
\end{align}
(In the following we set $2 \pi \alpha' = 1$.)
The Dirac quantization for this constrained system
has been carried out in Refs.%
\cite{KokadoEtAL2000},\cite{Chu2000}.
The result is summarized as follows:
First of all we go over to the Lorentz frame
in which the electric field is absent,
and then diagonalize the electromagnetic field tensor.
The total transformation is accomplished by the matrix
\begin{align}
  & S^\mu{}_\nu
  = \begin{pmatrix} a & - b/\sqrt{2} & - b/\sqrt{2} \\
                    0 & - i/\sqrt{2} &   i/\sqrt{2} \\
                  - b &   a/\sqrt{2} &   a/\sqrt{2} \end{pmatrix}~,
& & \left( S^{-1} \right)^\mu{}_\nu
  = \begin{pmatrix} a &             0 &            b \\
           b/\sqrt{2} &    i/\sqrt{2} &   a/\sqrt{2} \\
           b/\sqrt{2} &  - i/\sqrt{2} &   a/\sqrt{2} \end{pmatrix}~,
\label{eq:def-S}
\end{align}
to give
\begin{align}
  & \left( S^{-1} F S \right)^\mu{}_\nu
  = \begin{pmatrix} 0 &     0 & 0 \\
                    0 & i\, K & 0 \\
                    0 &     0 & - i\, K \end{pmatrix}~,
\label{eq:S_inv-F-S}
\end{align}
where $K = \sqrt{ B^2 - E^2 }$~(we assume $0 < E < B$) and
\begin{align}
  & a = \frac{B}{K}~, & & b = \frac{E}{K}~, & & a^2 - b^2 = 1~.
\end{align}
Then we have
\begin{align}
  & S^{-1} \begin{pmatrix} X^0 \\ X^1 \\ X^2 \end{pmatrix}
  = \begin{pmatrix} a\, X^0 + b\, X^2 \\
     \big( b\, X^0 + i\, X^1 + a\, X^2 \big)/\sqrt{2} \\
     \big( b\, X^0 - i\, X^1 + a\, X^2 \big)/\sqrt{2} \end{pmatrix}
  =: \begin{pmatrix} Z \\ X^{(+)} \\ X^{(-)} \end{pmatrix}~.
\label{eq:def-ZX}
\end{align}
The inverse relation is

\begin{align}
  & \begin{pmatrix} X^0 \\ X^1 \\ X^2 \end{pmatrix}
  = S \begin{pmatrix} Z \\ X^{(+)} \\ X^{(-)} \end{pmatrix}
  = \begin{pmatrix} a\, Z - b \big( X^{(+)} + X^{(-)} \big)/\sqrt{2} \\
                          - i \big( X^{(+)} - X^{(-)} \big)/\sqrt{2} \\
                - b\, Z + a\, \big( X^{(+)} + X^{(-)} \big)/\sqrt{2}
    \end{pmatrix}~.
\label{eq:rel-X-ZX}
\end{align}
The boundary condition Eq.(\ref{eq:bc}) become
\begin{align}
  & Z(\tau ,\sigma )\, \Big\vert_{\sigma=0, \pi} = 0~, \nonumber \\
  & \big( X'^{(\pm )} \pm i\, \rho K \dot X^{(\pm )} \big)
  \Big\vert_{\sigma=0, \pi} = 0~,
\label{eq:XpmX'} \\
  & X'^{i}(\tau ,\sigma ) \Big\vert_{\sigma=0, \pi}=0,
  \quad i=3,\cdots ,d-1
\nonumber 
\end{align}
Since $X^{(\pm)}$ couples with $K$,
their time derivatives have mode expansions like
\begin{align}
  & \dotX^{(\pm)}(\tau, \sigma)
  =  \sum_n\, e^{ -i (n \pm \omega) \tau}~
    \cos\big[~(n \pm \omega)
    \sigma \mp \pi \omega~\big]~\alpha^{(\pm)}_n~,
\label{eq:dotX-mode_decomposition}
\end{align}
which satisfy the boundary conditions (\ref{eq:XpmX'}),
 and $\tan \pi \omega := q\, K$.

On the other hand, $Z(\tau, \sigma)$ does not couple with any field,
so it is a free string, as well as other components,
$X^3,\cdot \cdot \cdot ,X^{d-1}$.
Their mode expansion are :
\begin{align}
  & Z(\tau, \sigma)
  = z + p^0\tau +i \sum_{n \ne 0}\frac{1}{n}\, e^{ -i n \tau}~
    \cos\big( n \sigma \big)~\alpha^{\ 0}_n ,
\label{eq:Z-mode_decomposition} \\
  & X^i(\tau, \sigma)
  = x^i + p^i\tau  + i \sum_{n \ne 0}\frac{1}{n}\, e^{ -i n \tau}~
    \cos\big( n \sigma \big)~\alpha^{\ i}_n, \quad i=3,\cdots ,d-1.
\label{eq:x-mode_decomposition}
\end{align}
The commutation relations for mode operators for $X^{(\pm )}$ are
\begin{align}
  & \big[\, \alpha^{(\pm)}_m\,,~\alpha^{(\mp)}_n~\big]
  = ( m \pm \omega  )~\delta_{m+n, 0}~,
\label{eq:CCR_alpha}
\end{align}
whereas for other free components, $Z,X^3,\cdot \cdot \cdot ,X^{d-2}$,
\begin{align}
  & \big[\, \alpha^{i}_m\,,~\alpha^{j}_n~\big]
  = \delta^{ij} m~\delta_{m+n, 0}~, \quad i,j=3,\cdots ,d-2.
\label{eq:CCR_alphaij} \\
  & \big[\, \alpha^{0}_m\,,~\alpha^{0}_n~\big]
  = -m~\delta_{m+n, 0}~.
\label{eq:CCR_alpha0}
\end{align}
We find that the noncommutativity of $X^{(\pm)}(\tau, \sigma)$
at $\sigma=0$ is
\begin{align}
  & \big[\, X^{(+)}(\tau, 0)\,,~X^{(-)}(\tau, 0)~\big] = \theta~,
& & \theta = \frac{q\, K}{ 1 + q^2\, K^2 }~.
\label{eq:CCR_X-def_theta}
\end{align}

The Virasoro operator is defined as
\begin{align}
  & L_n
  = \frac{1}{4} \int^{\pi}_{-\pi} d\sigma~e^{\pm i n \sigma}~
      : \big( \dotX \pm X' \big)^2 :~
  = \half~\sum_{l}: \alpha_l \cdot \alpha_{n-l} :~,
\label{eq:def-Virasoro_op}
\end{align}
where $A\cdot B=-A^0B^0 + A^{(+)}B^{(-)} + A^{(-)}B^{(+)} +A^3B^3
+\cdot \cdot \cdot +A^{d-1}B^{d-1} $,
and it satisfies the Virasoro algebra%
\footnote{The linear term in $\omega$ is necessary
in Eq.(\ref{eq:def-Virasoro_algebra}).
The original article \cite{KokadoEtAL2000}
misses the term, so it should be modified in this way.}
\begin{align}
  & \big[\, L_m\,,~L_n~\big]
  = ( m - n )\, L_{m+n}
  + \left\{~\frac{d-2}{12}\, m (m^2 - 1) - m\, \omega^2
     + m\, \omega~\right\}~\delta_{m+n, 0}~,
\label{eq:def-Virasoro_algebra}
\end{align}
The Hamiltonian corresponds to $L_0$,
and the Virasoro condition for physical state are
\begin{align}
  & L_n \ket{\psi} = 0~, \ \ n>0,
\label{eq:physicalstaten} \\
  & \left[ L_0 - \alpha (0) \right] \ket{\psi} = 0~.
\label{eq:physicalstate0}
\end{align}
where $\alpha(0)$ is Regge intercept.
The no-ghost problem will be discussed in Sec.\ref{sec:SGAstring}.
\section{The Hall conductivity} \label{sec:Hall_cond}
We consider especially the end-point of the string at $\sigma =0$. 
From Eq.(\ref{eq:rel-X-ZX}), $X^0(\tau, 0)$
and $X^2(\tau, 0)$ are given by
\begin{align}
  & X^0(\tau, 0) = a\, Z(\tau, 0) - b\, Y(\tau, 0)~,
\label{eq:X_zero}
\\
  & X^2(\tau, 0) = -b\, Z(\tau, 0) + a\, Y(\tau, 0)
                 = -\frac{b}{a}\, X^0(\tau, 0) + \frac{1}{a}\, Y(\tau, 0)~, 
\label{eq:x^2}
\end{align}
where
\begin{align}
 Y(\tau, \sigma) := ( X^{(+)} + X^{(-)} )/\sqrt{2}.
\label{eq:Y-def}
\end{align}
Now, for Eqs.(\ref{eq:X_zero}) and (\ref{eq:x^2}),
let us consider expectation values by stationary states
$\ket{\alpha ,p}$,
which are normalized by Dirac's delta function
as $\big\langle~p,\alpha ~\ket{\alpha ',p'}
=\delta _{\alpha , \alpha'}\delta (p-p')$,
where $p$ is the c.m. momentum of the string.
For an example, the expectation value of the momentum operator
$\Hat{p}$ is defined by the coefficient of the delta function
as $\bra{p,\alpha } \Hat{p} \ket{\alpha ',p'}= p \delta (p-p')$.
The expectation value for Eq.(\ref{eq:X_zero}) is
\begin{align}
   \bra{p,\alpha } X^0(\tau ,0) \ket{\alpha ,p'}
  := t \delta(p-p')
  = a p^0\tau \delta (p-p')
  - b \bra{p,\alpha } Y(\tau ,0) \ket{\alpha ,p'}~.
\label{eq:expectvalue}
\end{align}
Here we have set $\Exp{X^0(\tau ,0)}\sim t$,
which plays a role of the ordinary time.
The $\tau$ derivative of Eq.(\ref{eq:expectvalue}) becomes
\begin{align}
  \frac{\partial t}{\partial \tau } = a p^0~,
\label{eq:dtaudt}
\end{align}
because $\bra{\alpha } \dot{Y}(\tau ,0) \ket{\alpha }$ vanishes,
$\dot{Y}$  being linear in $\alpha _{n}^{(\pm)}$.

In the same way, the expectation value of Eq.(\ref{eq:x^2}) is
\begin{align}
   \bra{p,\alpha }X^2(\tau ,0)\ket{\alpha ,p'}
  = -\frac{b}{a}t\delta (p-p')
  + \frac{1}{a}\bra{p,\alpha }Y(\tau ,0)\ket{\alpha ,p'}~,
\label{eq:expectX2}
\end{align}
The $t$-derivative of this equation becomes
\begin{align}
   \frac{\partial }{\partial t}
    \bra{p,\alpha } X^2(\tau ,0) \ket{\alpha ,p'}
  &= - \frac{b}{a}\delta (p-p')
  + \frac{1}{a}\frac{\partial \tau}{\partial t }
     \bra{p,\alpha } \dot{Y}(\tau ,0) \ket{\alpha ,p'}
\nonumber \\
  &= - \frac{b}{a}\delta (p-p')~.
\label{eq:dX2/dt1}
\end{align}
Since the expectation value is defined by
the coefficient of the delta function,
we write Eq.(\ref{eq:dX2/dt1}) as
\begin{align}
   \frac{\partial }{\partial t}\Exp{X^2(\tau ,0)}
  = - \frac{b}{a} = - \frac{E}{B}~.
\label{eq:dX2/dt2}
\end{align}
Also
\begin{align}
   \frac{\partial }{\partial t}\Exp{X^1(\tau ,0)}
  = \frac{\partial \tau }{\partial t}
    \frac{\partial }{\partial \tau }\Exp{X^1(\tau ,0)}
  = \frac{1}{ap^0}\Exp{\dot{X}^1(\tau ,0)}=0~,
\label{eq:dX1/dt}
\end{align}
because the $\tau$-derivative of
$X^1 = - i \big( X^{(+)} - X^{(-)} \big)/\sqrt{2}$
is linear in $\alpha^{(\pm)}_n$.

The current density in the $i$-direction is given by
$J^i = q\, n\, (\partial/\partial t) \Exp{ X^i(\tau, 0) }$,
where $q\, n$ is the charge density.
Hence Eqs.(\ref{eq:dX2/dt2}) and (\ref{eq:dX1/dt}) mean
\begin{align}
  & J^2 = - \frac{q\, n}{B}\, E~,
\label{eq:J_two} \\
  & J^1 = 0~.
\label{eq:J_one}
\end{align}
This means that the Hall conductivity for charged strings
is $\sigma_{21} = - q\, n/B$,
and this value is of the same form as in the ordinary particle case.
\section{Spectcrum-generating algebra for the charged string}
\label{sec:SGAstring}
In this section we consider a spectrum-generating algebra (SGA)
for charged strings when antisymmetric background fields are present.
This is necessary to guarantee that the system is ghost-free.

First we recall SGA for a free string.
We define relevant operators as
\begin{align}
  & X^\mu (\tau )
  =  i \sum_{n \ne 0}\frac{1}{n}\, e^{ -i n \tau}~
    \alpha^{\mu }_n +x^\mu  + \tau p^\mu ~, 
\label{eq:X-mode_decomposition} \\
  & P^\mu (\tau )
  = \partial _\tau X^\mu (\tau )
  = \sum_{n }\, e^{ -i n \tau}~\alpha^{\mu }_n~,
\label{eq:P-mode_decomposition} \\
  & V(\tau ) = e^{iX^{-}(\tau )}~,
\label{eq:V-def} \\
  & A^{i}_n = \oint d\tau ~P^i(\tau )V(\tau )^n~;
\quad
    \oint d\tau = (2\pi )^{-1}\int ^{\pi }_{-\pi }~d\tau~,
\quad i=1,\cdots ,d-2,
\label{eq:A-def} \\
  & K_n = - \oint d\tau :\{~P^{+}(\tau )
  + \frac{1}{2}n^2P^{-}(\tau )\log P^{-}(\tau )\}V(\tau )^n~:~,
\label{eq:K-def}
\end{align}
where $P^{\pm}$,$X^{\pm}$ are light-cone variables defined by
$X^{\pm}=(X^0 \pm X^{d-1})/\sqrt{2}$, etc., with 
the c.m. variables $x^\mu $, $p^\mu $,
and the mode operators $\alpha ^\mu _n$ satisfying
\begin{align}
  & \big[\, x^\mu \,,~p^\nu ~\big]=i\eta ^{\mu \nu}.
\label{eq:CCR_xp} \\
  & \big[\, \alpha^{\mu }_m\,,~\alpha^{\nu }_n~\big]
  = \eta ^{\mu \nu}m~\delta_{m+n, 0}~,
  \quad (\mu ,\nu =0,1,\cdots ,d-1)~.
\label{eq:CCR_alphamn}
\end{align}
We work in the subspace with $p^- = 1$, so that
\begin{align}
 & \oint d\tau ~P^i(\tau )V(\tau )^n=\delta _{n,0}~.
\label{eq:A-V}
\end{align}

SGA is summarized as follows:
\begin{align}
  & \big[\, A^i_m \,,~A^j_n ~\big]
  = m\delta  ^{ij}\delta _{m+n,0}~, \quad i,j=1,\cdots ,d-2
\nonumber \\
  & \big[\, A^{i }_m\,,~K_n~\big] = m~A_{m+n}^{i}~.
\label{eq:CCR_AmKn} \\
  & \big[\, K_m\,,~K_n~\big]
  = (m-n)~K_{m+n} + 2m^3\delta _{m+n,0}~.
\nonumber
\end{align}
The spectrum-generating operators $A^i_n$, $K_n$
are commutable with the Virasoro operator $L^{(0)}_n$
for the free string, and generate all the physical states
satisfying the Virasoro conditions.
The commutation relations (\ref{eq:CCR_AmKn}) suggest an isomorphism
\begin{align}
  & A^i_n \sim \alpha ^i_n~, \quad K_n \sim L^{(0) T}_n~,
\label{eq:isomorphism}
\end{align}
where $L^{(0) T}_n$ is the transverse part of the Virasoro operator
\begin{align}
   & L^{(0) T}_n := -a\delta _{n,0} + \frac{1}{2}\sum_{i=1}^{d-2}\,
   \sum_{l}\, :\alpha^i_{n-l} \alpha^i_{l}: ~, 
\label{eq:T-Virasoro-def} 
\end{align}
satisfying
\begin{align}
  & \big[\, \alpha ^i_m \,,~\alpha ^j_n ~\big]
  = m\delta  ^{ij}\delta _{m+n,0}~, \quad i,j=1,\cdots ,d-2
\nonumber \\
  & \big[\, \alpha ^{i }_m\,,~L^{(0) T}_n~\big]
  = m~\alpha ^{i }_{m+n}~,
\label{eq:CCR_AmKn-II} \\
  & \big[\, L^{(0) T}_m\,,~L^{(0) T}_n~\big]
  = (m-n)~L^{(0) T}_{m+n}
  + \big[\,\frac{d-2}{12}(m^3-m) + 2ma~\big]\delta _{m+n,0}~.
 \nonumber
\end{align}
The c-number $a$ is Regge intercept $a=\alpha (0)$.
Comparing Eqs.(\ref{eq:CCR_AmKn}) with (\ref{eq:CCR_AmKn-II}),
we find that the isomorphism is completed if $d=26$ and $a=1$.

Now let us return to the charged string
in Sec.\ref{sec:Hall_cond}.%
\footnote{The discussion in this section is valid also
for a charged string with charges at both ends.
The parameter $\omega $ is defined by $\omega =\omega_0 - \omega_\pi$
with $\tan \pi \omega _0 = \rho K$ and
$\tan \pi \omega _\pi = \rho(\pi) K$.
In that case the $\omega$ in Eq.(\ref{eq:d_alpha-condition})
should be replaced by $|\omega|$.} We consider the string coordinates,
$Z, X^{(+)}, X^{(-)}, X^3, \dots, X^{d-2},X^{d-1}$.
The light-cone variables are now $X^{\pm}:=(Z \pm X^{d-1})/\sqrt{2}$
(which should be distinguished from $X^{(\pm)}$).

For the $(\pm)$ components, instead of $P^\mu (\tau )$
in Eq.(\ref{eq:P-mode_decomposition}) we adopt
\begin{align}
  P^{(\pm)}(\tau )= \sum_{n}\,e^{-i(n\pm \omega )\tau }\alpha ^{(\pm)}_n~,
\label{eq:P_mode_exp}
\end{align}
which satisfy
\begin{align}
  & \big[\, P^{(\pm)}(\tau )\,,~P^{(\mp)}(\tau' )\big]
  = \sum_{n}\,(n \pm \omega )e^{-i(n\pm \omega )(\tau - \tau')}
  = i\partial _\tau \delta _{\pm\omega }\delta (\tau - \tau')~.
\label{eq:CC_PP}
\end{align}
Here $\delta _{\omega }(s)$ is defined
as $\delta _{\omega }(s)=\exp(-i\omega s)\delta(s)$
with 2$\pi$-period delta function $\delta(s)$,
and has a property, $\delta _{\omega }(s+2\pi )
=\exp(-2\pi i\omega)\delta_{\omega }(s)$.

Corresponding to $A^{i}_n$ in Eq.(\ref{eq:A-def})
we define the SGA operators for the $(\pm)$ components as
\begin{align}
 & A^{(\pm )}_n=\oint d\tau ~P^{(\pm )}(\tau )V(\tau )^{n\pm \omega }~.
\label{eq:Apm=def}
\end{align}
For the sake of $p^{-1}=1$, $V(\tau )$ carries
a factor $e^{i\tau }$ and hence $V(\tau )^{n \pm \omega}$
a factor $e^{i(n\pm \omega )\tau }$.
The non-periodic factor $e^{i\omega \tau }$
is canceled by the similar factor in  $P^{\pm}(\tau)$.
In the following calculations this cancellation
always occurs among $\delta _{\pm \omega }, V(\tau )^{n \pm \omega}$
and $P^{(\pm)}$.
 
We are to show the commutation relations
\begin{align}
  & \big[\, A^{(\pm)}_m \,,~A^{(\mp)}_n ~\big]
  =(m \pm \omega )\delta _{m+n,0}~,
\label{eq:CCR_ApmAmp2} \\
  & \big[\, A^{(\pm )}_m\,,~K_n~\big] = (m\pm \omega )~A^{(\pm)}_{m+n}~.
\label{eq:CCR_AmKn-III}
\end{align}
which guarantee the $(\pm)$ parts of the isomorphism
$A^{(\pm)}_n \sim \alpha ^{(\pm)}_n$, and to show also that
the $A^{(\pm)}_n$'s commute with the Virasoro operators.

The commutator  $\big[\, A^{(\pm)}_m \,,~A^{(\mp)}_n ~\big]$
is calculated as
\begin{align}
   \big[\, A^{(\pm)}_m \,,~A^{(\mp)}_n ~\big]
  &= \oint d\tau \oint d\tau '
     i\partial _\tau \delta _{\pm \omega }(\tau -\tau ')
     V(\tau )^{m\pm \omega }V(\tau ')^{n\mp \omega }
\nonumber \\
  &= \oint d\tau V(\tau )^{m\pm \omega }
  i\partial_\tau V(\tau )^{n\mp \omega }
  = -(n\mp \omega )\oint \tau P^{-}V(\tau )^{m+n}
\label{eq:CCR_ApmAmp3}
\\
  &= -(n \mp \omega )\delta _{m+n,0}~, \nonumber 
\end{align}
giving Eq.(\ref{eq:CCR_ApmAmp2}).
In the same way we have
\begin{align}
   \big[\, A^{(\pm )}_m\,,~K_n~\big]
  &= - \oint d\tau \oint d\tau ' P^{(\pm)}(\tau )
       \big[\, V(\tau )^{m\pm \omega }\,,~P^{+}(\tau ')~\big]
       V(\tau ')^{n}
\nonumber \\
  &= \oint d\tau \oint d\tau ' P^{(\pm)}(\tau )
     V(\tau )^{m\pm \omega }(m\pm\omega )
     \delta (\tau -\tau ')V(\tau ')^{n}
\nonumber \\
  &= (m\pm\omega )\oint d\tau P^{(\pm)}(\tau )
     V(\tau )^{m + n \pm \omega }~,
\label{eq:CCR_Am+-Kn}
\end{align}
to show Eq.(\ref{eq:CCR_AmKn-III}).

The commutativity of $A^{(\pm)}_n$ with the Virasoro operator
$L_n$ (\ref{eq:def-Virasoro_op}) with $F \neq 0$
can be shown as follows: We define $L(\tau)$ by
\begin{align}
  & L(\tau ) = \sum_{n}e^{-in\tau }L_{n}
  =: P^{+}(\tau )P^{-}(\tau ): + :P^{(+)}(\tau )P^{(-)}(\tau ):
  + \frac{1}{2}\sum_{i=3}^{d-2}\,:\{P^i(\tau )\}^{2}: ~.
\label{eq:L-exp}
\end{align}
In order to show  $\big[\, L(\tau ) \,,~A^{(\pm)}_n ~\big]=0$,
we first calculate the first term of $L(\tau )$  as
\begin{align}
   \big[\, :P^{+}(\tau )P^{-}(\tau ):\,,~A^{(\pm)}_n ~\big]
  &= \oint d\tau' P^{-}(\tau )
     \big[\, P^{+}(\tau)\,,~V(\tau ')^{n\pm \omega }~\big]
     P^{(\pm )}(\tau ')
\nonumber \\
  &= \oint d\tau ' P^{-}(\tau )(n\pm \omega )
     \delta (\tau - \tau')V(\tau ')^{n\pm \omega }
     P^{(\pm)}(\tau ')
\nonumber \\
  &= (n \pm \omega )P^{(\pm)}(\tau )P^{-}(\tau )
     V(\tau )^{n\pm \omega }~.
\label{eq:CCR_PPAn}
\end{align}
For the second term we have
\begin{align}
   \big[\, :P^{(+)}(\tau )P^{(-)}(\tau ):\,,~A^{(\pm)}_n~\big]
  &= \oint d\tau ' P^{(\pm)}(\tau )\big[\, P^{(\mp)}(\tau)\,,~
     P^{(\pm)}(\tau ')~\big]V(\tau ')^{n\pm \omega }
\nonumber \\
  &= \oint d\tau ' P^{(\pm)}(\tau )i\partial_{\tau }
     \delta _{\mp\omega }(\tau - \tau')V(\tau ')^{n\pm \omega }
\nonumber \\
  &= P^{(\pm)}(\tau )i\partial _{\tau }V(\tau )^{n\pm \omega }
  = -(n\pm\omega )P^{(\pm)}(\tau )P^{-}(\tau )V(\tau )^{n\pm\omega }~.
\label{eq:CCR_PPAn-second}
\end{align}
Since the third term is irrelevant, we get
\begin{align}
  & \big[\, L(\tau ) \,,~A^{(\pm)}_n ~\big]=0~.
\label{eq:CCL_Apmn}
\end{align}

To sum up, the modified SGA for the $(\pm)$-components is given by
\begin{align}
  & \big[\, A^{(\pm)}_m \,,~A^{(\mp)}_n ~\big]
  = (m \pm \omega )\delta _{m+n,0}~, 
\nonumber \\
  & \big[\, A^{(\pm)}_m\,,~K_n~\big]
  = (m \pm \omega )~A^{(\pm)}_{m+n}~,
\label{eq:CCR_AmKn-IV} \\
  & \big[\, K_m\,,~K_n~\big]
  = (m-n)~K_{m+n} + 2m^3\delta _{m+n,0}~.
\nonumber
\end{align}
These are to be compared with the commutation relations
\begin{align}
  & \big[\, \alpha ^{(\pm)}_m \,,~\alpha ^{(\mp)}_n ~\big]
  = (m \pm \omega )\delta _{m+n,0}~, 
\nonumber \\
  & \big[\, \alpha ^{(\pm)}_m\,,~L_n^T~\big]
  = (m \pm \omega )~\alpha ^{(\pm)}_{m+n}~,
\label{eq:CCR_alphamLn} \\
  & \big[\, L_m^T\,,~L_n^T~\big]
  = (m-n)~L_{m+n}^T + [\frac{d-2}{12}(m^3-m)
  - m(\omega ^2-\omega -2a)]\delta _{m+n,0}~.
\nonumber
\end{align}
where
\begin{align}
  & L_{n}^{T} := -a\delta _{n,0}
  + \frac{1}{2}\sum_{l}:[\alpha ^{(+)}_{n-l}\alpha ^{(-)}_{l}
  + \alpha ^{(-)}_{n-l}\alpha^{(+)}_{l} +\alpha ^3_{n-l}\alpha ^3_{l}
  + \cdots +\alpha ^{d-2}_{n-l}\alpha ^{d-2}_{l}]:~.
\label{eq:L-exp-II}
\end{align}
Comparing Eqs.(\ref{eq:CCR_AmKn-IV}) with (\ref{eq:CCR_alphamLn}),
the isomorphism
\begin{align}
  & A_n^{(\pm)} \sim \alpha _n^{(\pm)}~, \quad K_n \sim L_n^{T},
\label{eq:isomorphism2}
\end{align}
is completed if
\begin{align}
   d=26~, \quad
   a :=\alpha (0)=1-\frac{\omega }{2} + \frac{\omega ^2}{2}~.
\label{eq:d_alpha-condition}
\end{align}
The other transverse components obey the same algebra
as in the free case, and we have the same isomorphism
as Eqs.(\ref{eq:isomorphism}).
Once we have SGA, then we can conclude that our system
is ghost-free provided the conditions (\ref{eq:d_alpha-condition})
are satisfied.
\section{Concluding remarks}
We have calculated the Hall conductivity for the charged strings,
which end on the D2-brane with a constant background magnetic field
perpendicular to the brane.
The result is given by Eq.(\ref{eq:J_two}), $\sigma_{21} = - q\, n/B$,
which coincides completely with
that of the ordinary two-dimensional electron system.
In this derivation the mode expansion
(\ref{eq:dotX-mode_decomposition}) of $\dotX^{(\pm)}$
has been essential, where the commutation relation (\ref{eq:CCR_alpha})
for mode operators $\alpha^{(\pm)}_n$
is related with the noncommutativity
between $X^{(+)}(\tau, 0)$ and $X^{(-)}(\tau, 0)$.
However, the noncommutativity itself has not played any role
in this derivation.

When both ends of the string have charges,
i.e. $\rho (0)=q$, $\rho(\pi )=-q'$, 
and they attach to the D2-brane,
we have also the same result  $\sigma _{21}=\sigma -(q+q')n/B$.
It may be intetesting problem to consider a case where open string end on
two different D-branes with different world volume field strengths.

  As another main result,
we have constructed SGA for charged strings with constant antisymmetric
background fields.
We conclude that our system is ghost-free
if the space-time dimension is $d=26$ and the Regge intercept
is $\alpha (0)=1-\omega /2 + \omega ^2/2$,
where $\omega $ is the cyclotron frequency of the charged string. 

Any DDF state constructed of the modified spectrum-generating operators
eventually did not contribute to the Hall conductivity,
but only the ground state has been remained. 
Generally speaking, however,
the modified SGA will be crucially important
in the theory of charged strings. 

Finally we recall that,
in order to obtain the quantum Hall effect for electrons, 
it is important to take into account of effects of impurity potentials
or electron-electron scatterings. 
Otherwise, we get only the pure result (\ref{eq:def-class_Hall_cond}).
The same thing may happen to the charged strings. 
We expect also the quantum Hall effect for the charged strings,
if we take into account of effects of impurity potentials
or string-string scatterings.
This interesting problem will be remained in future studies. 
\begin{acknowledgments}
It is a pleasure to thank T. Okamura for continuous discussions.
\end{acknowledgments}

\end{document}